\documentclass[aps,prl,showpacs,twocolumn]{revtex4-1}

\usepackage{amsmath}
\usepackage{graphicx}
\usepackage{dcolumn}
\usepackage{bm}
\usepackage{color}
\usepackage{soul}

\def\Re{\mathop{\rm Re}\nolimits}
\def\Im{\mathop{\rm Im}\nolimits}

\begin{document}
\title{Auto-modulation versus breathers in the nonlinear stage of modulational instability
}

\author{Matteo Conforti$^{1*}$, Sitai Li$^2$, Gino Biondini$^2$, and Stefano Trillo$^3$}

\affiliation{$^1$Univ. Lille, CNRS, UMR 8523-PhLAM-Physique des Lasers Atomes et Mol\'ecules, F-59000 Lille, France\\
$^2$Department of Mathematics, State University of New York at Buffalo, New York 14260, USA\\
$^3$Department of Engineering, University of Ferrara, Via Saragat 1, 44122 Ferrara, Italy\\
Corresponding author: matteo.conforti@univ-lille.fr}




\begin{abstract}
The nonlinear stage of modulational instability in optical fibers induced by a wide and easily accessible class of localized perturbations is studied using the nonlinear Schr\"odinger equation.
It is showed that the development of associated spatio-temporal patterns is strongly affected by the shape and the parameters of the perturbation.
Different scenarios are presented that involve an auto-modulation developing in a characteristic wedge, possibly coexisting with breathers which lie inside or outside the wedge. 
\end{abstract}


\maketitle
Modulational instability (MI) is an ubiquitous nonlinear process that entails the growth of low frequency perturbations on top of a strong continuous wave (CW) pump \cite{ZO09}.
Optical fibers are an ideal ground to investigate the nonlinear stage of MI \cite{Kibler10,Kibler12,Dudley14,optimalMI,Mussot18}, i.e. the regime of growth saturation, which recently attracted a strong interest due also to its connection to rogue wave formation \cite{Dudley14}, breather solitons \cite{Kibler10,Kibler12}, and recurrence phenomena \cite{Kibler10,Mussot18}, all suitably described in the framework of the nonlinear Schr\"odinger equation (NLSE). While the research was mostly focused on the dynamics of periodic perturbations \cite{Kibler10,optimalMI,Mussot18}, MI induced by localized perturbations is still controversial. 

To date, two main approaches have been proposed to study the nonlinear stage of MI. In the first one, nonlinear MI gives rise to the onset of spontaneous oscillations (auto-modulation) that expand in time over a characteristic spatio-temporal wedge, smoothly connecting to the CW outside such wedge \cite{K67,KK68,El93,BM16,BLM16}. This scenario was first studied in \cite{K67,KK68} and later described in terms of Whitham modulation theory in \cite{El93}. Only recently, however, it was finally put on rigorous ground through asymptotic theory based on the inverse scattering transform (IST) associated with the NLSE \cite{BK14,BM16,BLM16,BM17}. As was shown in \cite{kuznetsov,BF15}, the phenomenon is driven by the continuum spectrum in the IST problem, and is unrelated to {\em breathers} (i.e., breathing solitons on finite background, associated with discrete IST spectrum). Moreover, it is rather universal, being independent of the specific perturbations or the integrable nature of the NLSE, and arising instead for a broader class of dynamical models \cite{BLMT17}. 
The phenomenon has been recently observed in fiber experiments \cite{Kraych18} and is also closely linked to other oscillating structures observed from an evolving step in power 
\cite{Audo18}.

In the second approach, nonlinear MI was described in terms of particular pairs of breathers with opposite velocities, termed super-regular breathers (SRB) \cite{ZG13,GZ14}, which superpose in input in such a way to yield a sufficiently small oscillating perturbation of the CW. The observation of such type of breather pairs has been recently reported both in optics and hydrodynamics \cite{Kibler15}. It must be noticed, however, that in both experiments the input was designed to carefully fit the initial theoretical datum for the breathers. Conversely, their excitability under sufficiently generic perturbations is largely unknown, though specific cases has been recently discussed \cite{Kibler17,Gelash18,Zhao16}.

Here, our aim is to reconcile the two approaches and show that, more generally, auto-modulations and breather pair formation coexist. Importantly, we show that, while the wedge velocities are fixed only by the CW, the breather velocities depend on the perturbation and hence can set the pair outside or inside the wedge. We also show that the excitation of breathers requires a proper decay (exponential) of the perturbation envelope. Perturbations that decay faster give rise only to specific auto-modulations, and provide an explanation for this phenomenon.  These results are of crucial importance for designing further fiber experiment and for full understanding the complex dynamics driven by nonlinear stage of MI.

We start with the NLSE conveniently written in the form
\begin{equation}\label{nls}
i q_z + q_{tt} + 2(|q|^2-q_0^2) q = 0,
\end{equation}
where the complex envelope of the real-world electric field, distance, and time read respectively as $E(Z,T)=q(z,t) \sqrt{P} \exp(i \gamma P Z)$, $Z=z (2 Z_{nl})$ and $T=t \sqrt{|\beta_2| Z_{nl}}$. Here $Z_{nl}=(\gamma P)^{-1}$ is the nonlinear length associated with reference power $P$, $\gamma$ is the fiber nonlinear coefficient, and $\beta_2<0$ is the dispersion.
We consider Eq.~(\ref{nls}) subject to the boundary conditions $q(t=\pm\infty)=q_{\pm}$ where $|q_{\pm}|=q_0$ is the normalized CW background (in all examples we take $q_0=1$, which implies $P$ to be the CW pump power). In particular, we are interested to describe the distinctive nature of the nonlinear MI evolutions that develop from sufficiently generic localized perturbation of the CW. To this end, we report results of the numerical integration of the NLSE (\ref{nls}) obtained with standard split-step method, and initial conditions
\begin{equation}\label{IV}
q(z=0,t)=q_0+ a~e^{i\phi} f_{p}(t/t_0) \cos[\omega (t - \overline{t})],
\end{equation}
where we take the perturbation envelope $f_p$ to have either Gaussian shape $f_{p}=\exp(-t^2/t_0^2)$ or hyperbolic secant (sech) shape $f_{p}={\rm sech}(t/t_0)$, $t_0$ denoting the width. 
This choice of even envelopes $f_p$ is made for sake of simplicity and lead to symmetric breather pairs (the asymmetric case will be analysed elsewhere). Without loss of generality we present cases with $\overline{t}=0$, since $\overline{t}$ is found to affect only the phase of the internal breathing of solitons and not the overall dynamics.

For a given initial condition, we assess the presence of breathers and their properties on the basis of the IST with non-zero boundary conditions \cite{BK14,BM16}.
To this end we search numerically for discrete eigenvalues of the following Zakharov-Shabat scattering problem associated with Eq.~(\ref{nls}) \cite{BM16}
\begin{equation}\label{scattering}
\phi_t = M\,\phi,\qquad 
M = \begin{pmatrix}
ik & q\\
-q^* &-ik
\end{pmatrix}
\end{equation}
where $k$ is the spectral parameter, $\phi=(\phi_1,\phi_2)$ is a matrix solution, and $\phi_{1,2}$ are column vectors. Since the spectral data are independent of $z$, it is sufficient to compute them at $z=0$. Thus, we search for discrete eigenvalues by letting $q=q(z=0,t)$ [Eq. (\ref{IV})]. 
The Jost functions $\phi_{\pm}$ are solutions of Eq.~(\ref{scattering}) whose columns tend to pure Fourier modes as $t\to-\infty$ or as $t\to\infty$, respectively, with a temporal dependence of the type 
$e^{\pm i \lambda t}$ and frequency given by 
$\lambda(k)=\sqrt{\smash{k^2+q_0^2}\phantom{|}}$ (see \cite{BK14} for details).
The two sets of Jost solutions are not independent, and are related by he scattering matrix $S(k)$ via the scattering relation $\phi_-(t,z,k)=\phi_+(t,z,k)S(k)$. 
The zeros of the element $s_{22}(k)$ of $S$ define the discrete spectrum and give the soliton content of the initial condition.

To calculate the scattering data numerically, we fix a time window $ [-\overline{T},\overline{T}]$, outside of which the potential is taken constant ($q=q_0$). For a given $k$, we fix $\phi_+(\overline{T},k)$ as initial condition of Eq.~(\ref{scattering}) and integrate backwards from $\overline{T}$ to $-\overline{T}$, to get $\phi_+(-\overline{T},k)$. Integration is performed semi-analytically, by splitting the interval in $N$ parts and assuming a constant potential over each part, as in \cite{Boff1992}. 
The solution can be written as the product of $N$ matrices applied to the initial condition. The other Jost function $\phi_-(-\overline{T},k)$ is known, and this allow to find the scattering matrix, and in particular $s_{22}(k)=\det(\phi_{+1}(-\overline{T},k),\phi_{-2}(-\overline{T},k))/d(k)$, where $d(k)=2\lambda/(k+\lambda)$ \cite{BM17}. We both map $s_{22}(k)$ on a grid in the complex $k$-plane, and use a root-finding algorithm to find numerically the zeros of $s_{22}(k)$ in the complex plane.

\begin{figure}[t!]
\centerline{\includegraphics[width=\linewidth]{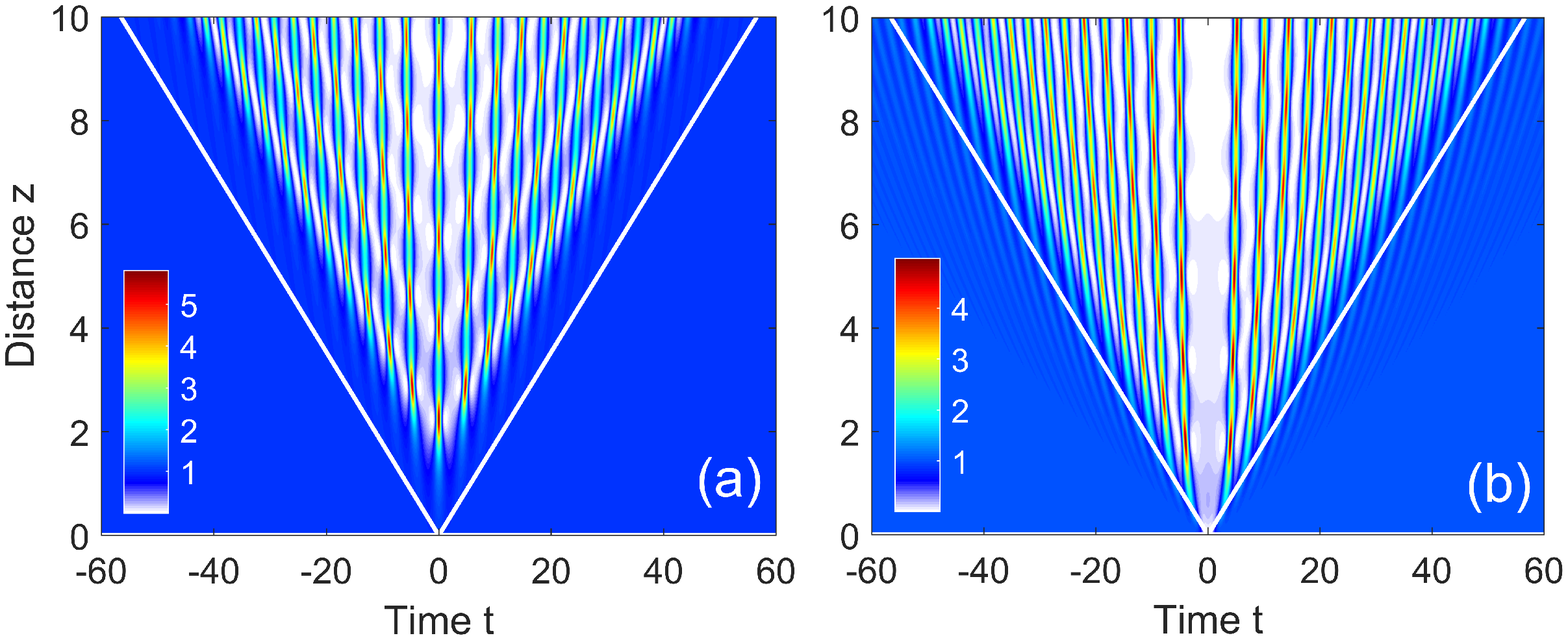}}
\caption{False color plot of power $|q|^2$ from numerical solution of NLSE (\ref{nls}). Here the CW $q_0=1$ is perturbed as in Eq. (\ref{IV}), Gaussian shape, $\omega=0$ and parameters:
(a) $a=0.1$, $\phi=\pi/2$, $t_0=1$; (b) $a=1$, $\phi=\pi$, $t_0=1$. White lines indicate the asymptotic wedge velocities $\pm V_w$ (slopes $t=\pm 4\sqrt{2} z$).}
\label{fig:wedge}
\bigskip
\centerline{\includegraphics[width=\linewidth]{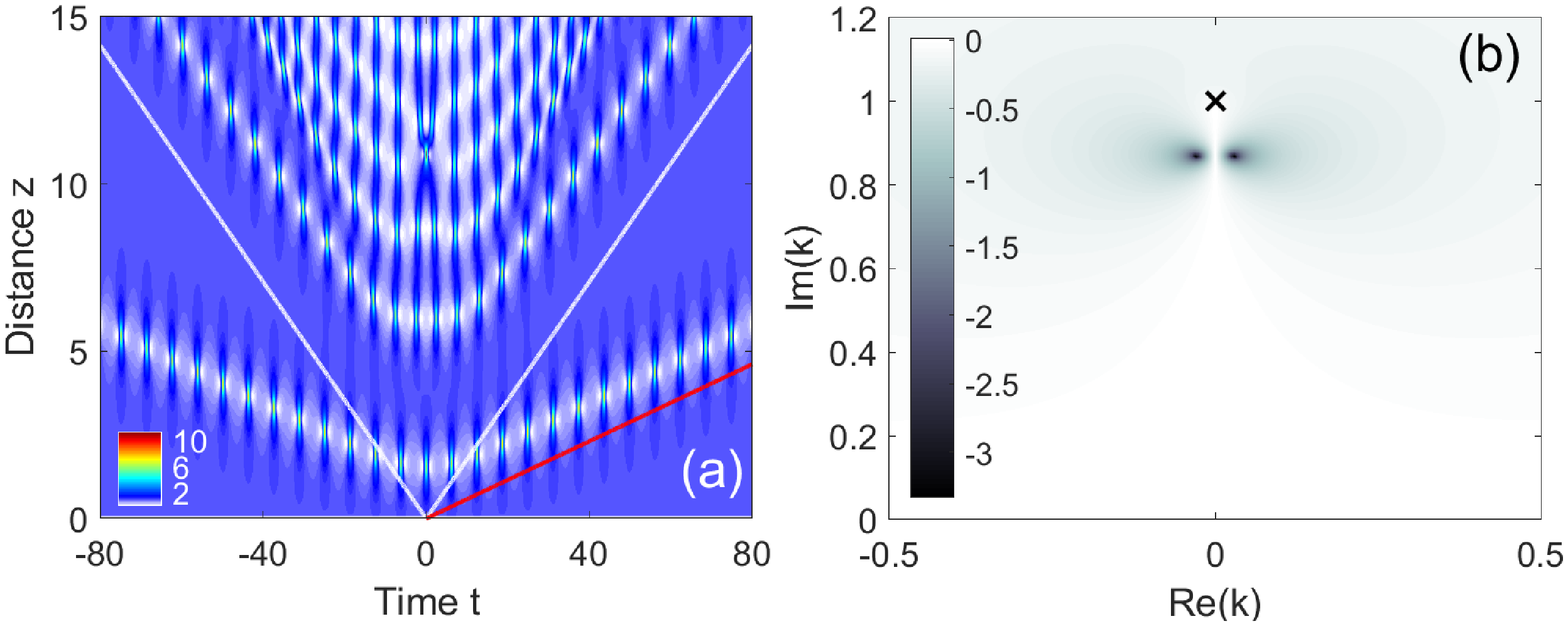}}
\caption{(a) Numerical solution of NLSE (\ref{nls}) with input (\ref{IV}), sech-shaped perturbation of the CW $q_0=1$,  with parameters:
$a=0.2$, $\phi=\pi/2$, $t_0=10$, $\omega=1$. Red solid line indicates the velocity $V_s$ of the right breather. 
(b) IST analysis of initial condition: false color plot of $\log(|s_{22}(k)|)$ in the complex plane, showing
a pair discrete eigenvalues found at 
$k=\pm 0.0285+0.870i$ ($R=1.059$, $\alpha=0.519$), 
giving a soliton velocity $V_{s}=17.35$. The cross marks $k=1i$ (Peregrine soliton).}
\label{fig:SRB_ext}
\end{figure}

When the initial condition contains no discrete eigenvalues, MI gives rise to a non-stationary auto-modulation, that is, a slow modulation of the oscillating cn-oidal wave solution of the NLSE. Such modulation spontaneously develops inside a characteristic wedge-shaped region in the $(t,z)$ plane (see Fig.~\ref{fig:wedge}), delimited by asymptotic slopes $dt/dz=\pm V_w$, where $V_w=4\sqrt{2} q_0={}${}$\mathrm{max}|dk/d\omega|$ \cite{El93,BM16,Kraych18} (the maximum occurs at $\omega=\sqrt{6}q_0$;  note also that hereafter we denote as velocities $V=dt/dz$ as usual in soliton theory, though, strictly speaking, these would be inverse velocities). The velocity $V_w$ is physically interpretable as the inverse linear group-velocity $(dk/d\omega)$ of the slowest components  that move away from the initial perturbation \cite{El93,BLMT17}. Along such edges the amplitude of the oscillations tends to vanish, smoothly connecting to the CW, which remains unperturbed outside the wedge. In Fig.~\ref{fig:wedge} we contrast two different wedge-shaped evolutions arising both from Gaussian perturbations at zero frequency ($\omega=0$), though with different amplitude $a$ and phase $\phi$. In Fig.~\ref{fig:wedge}(a) the perturbation is weak $(a=0.1)$ causing the auto-modulation to significantly develop only after a finite distance $z \simeq 1$, while the phase $\phi=\pi/2$ (though similar pattern is obtained for $\phi=0$) is such that  a central peak is present in $t=0$, which can be regarded locally in time as a soliton (the modulus of the Jacobian function that describes the oscillating pattern tends to one in $t=0$, as it is the case for solitons). Conversely, a negative perturbation ($\phi=\pi$) gives rise to central dip as shown in Fig.~\ref{fig:wedge}(b) \cite{Kraych18}, which also show that the structure originate at $z \simeq 0$ owing to the stronger amplitude ($a=1$) of the perturbation.

More generally, however, the auto-modulation in the wedge can coexist with breather pairs, whenever the initial condition turns out to contain discrete IST eigenvalues. 
A typical case corresponding in Eq.~(\ref{IV}) to the unstable frequency $\omega=1$ and a weak ($a=0.2$) and wide ($t_0=10$) sech-shaped envelope is reported in Fig.~\ref{fig:SRB_ext}. The numerical integration of the NLSE [see Fig.~\ref{fig:SRB_ext}(a)] shows that, while an auto-modulation still develops inside the wedge at finite distance, a pair of symmetric breathers, namely SRBs according to Refs. \cite{ZG13,GZ14,Kibler15}, clearly emerge since the early stage. The SRBs propagate with opposite velocities $\pm V_s$, and are {\em fast} compared with the asymptotic wedge velocity (i.e. $V_s>V_w$), thus propagating externally to the wedge. This is also supported by the outcome of our IST analysis of Eq.~(\ref{scattering}), displayed in Fig.~\ref{fig:SRB_ext}(b), where we show $\log |s_{22}|$ in the complex plane.  The two deep minima in Fig.~\ref{fig:SRB_ext}(b) constitute a good numerical approximation of a pair of eigenvalues (zeros of $s_{22}$), which we find at $k=\pm 0.0285+0.870i$. Their symmetric location around the imaginary axis indicates that the two breathers are identical except for their opposite velocities, given by the expression \cite{BK14}
\begin{equation} \label{vel}
V_s = 2 \left(\Re(k) + \Im(k)\frac{\Re[\lambda(k)]}{\Im[\lambda(k)]} \right).
\end{equation}
Equation (\ref{vel}) gives, for Fig.~\ref{fig:SRB_ext}(a), $V_{s}=\pm 17.35$, which are fully consistent with the simulation (see red line).
We refer the interested reader to Refs.~\cite{BK14,ZG13,GZ14,Kibler15} for explicit expressions for these breathers. 
The eigenvalue in \cite{ZG13,GZ14,Kibler15} is given in polar form through the parameters $(R,\alpha)$, which are easily linked to our parameters $(\Re(k), \Im(k))$ as $\Re(k)=\frac{1}{2}\left(R-R^{-1}\right)\sin\alpha$ and $\Im(k)=\frac{1}{2}\left(R+R^{-1}\right)\cos\alpha$, whereas the velocity in Eq.~(\ref{vel}) can be cast in the form 
$V_s=2[(R^4+1)/R(R^2-1)]\sin\alpha$.


Generally speaking, we have found that SRBs such as those in Fig.~\ref{fig:SRB_ext} emerge only from sech-shaped perturbations. This can be explained by an insightful (though heuristic) argument based on the MI amplification process. We recall that purely periodic modulations grow exponentially until temporal peaks are formed at a characteristic distance, beyond which the power flow reverses \cite{Dudley14,optimalMI}. When a localized envelope weighs the modulation as in Eq.~(\ref{IV}), these peaks are expected to emerge at non-uniform distances owing to the local weight (in time) of the perturbation. Indeed we can assume that the perturbation generally grows like $f_p(t) \exp(g z)$, where $g=g(\omega)=\omega \sqrt{4-\omega^2}$ is the MI gain. 
When this argument is specialized to $f_p(t)={\rm sech}(t/t_0)$, the growth over the tails proceeds approximately as $\exp[g(\omega) z \pm t/t_0]$. The peaks emerge for a uniform growth, i.e. constant argument $g(\omega) z \pm t/t_0$, which implies a distance $z$ that scales linearly with time $t$, with $V_s=dt/dz=\pm g(\omega) t_0$ giving a reasonable approximation of the velocities of the pair. 
($V_s \simeq 17.32$ for Fig.~\ref{fig:SRB_ext}). 
In other words the breathers are sustained by the usual MI amplification process, with their constant velocities being intimately related to the correct (exponential) decay of the perturbation envelope.

\begin{figure}[t!] 
\centering
\includegraphics[width=\linewidth]{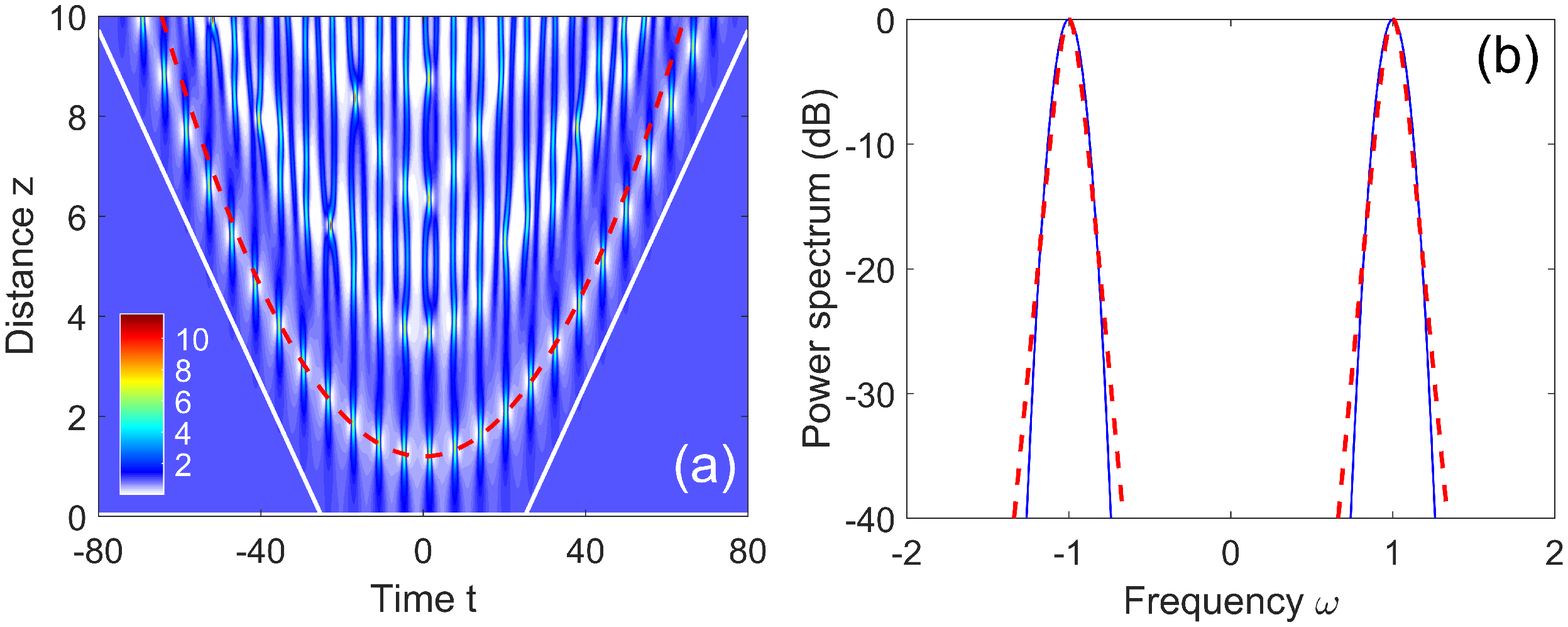}
\caption{(a) As in Fig.~\ref{fig:SRB_ext}(a), input Gaussian envelope, $a=0.2$, $\phi=0$, $t_0=16.5$, $\omega=1$. 
Red dashed curve indicates locus of peaks due to MI amplification: $g(\omega)z - t^2/t_0^2 = const.$ (see text; here constant used as best-fit parameter). 
(b) Input power spectrum (log scale, CW suppressed for clarity) for Gaussian (solid blue curve) and sech-shape of Fig.~\ref{fig:SRB_ext} (dashed red curve).}
\label{fig:gaus}
\bigskip
\centering
\includegraphics[width=\linewidth]{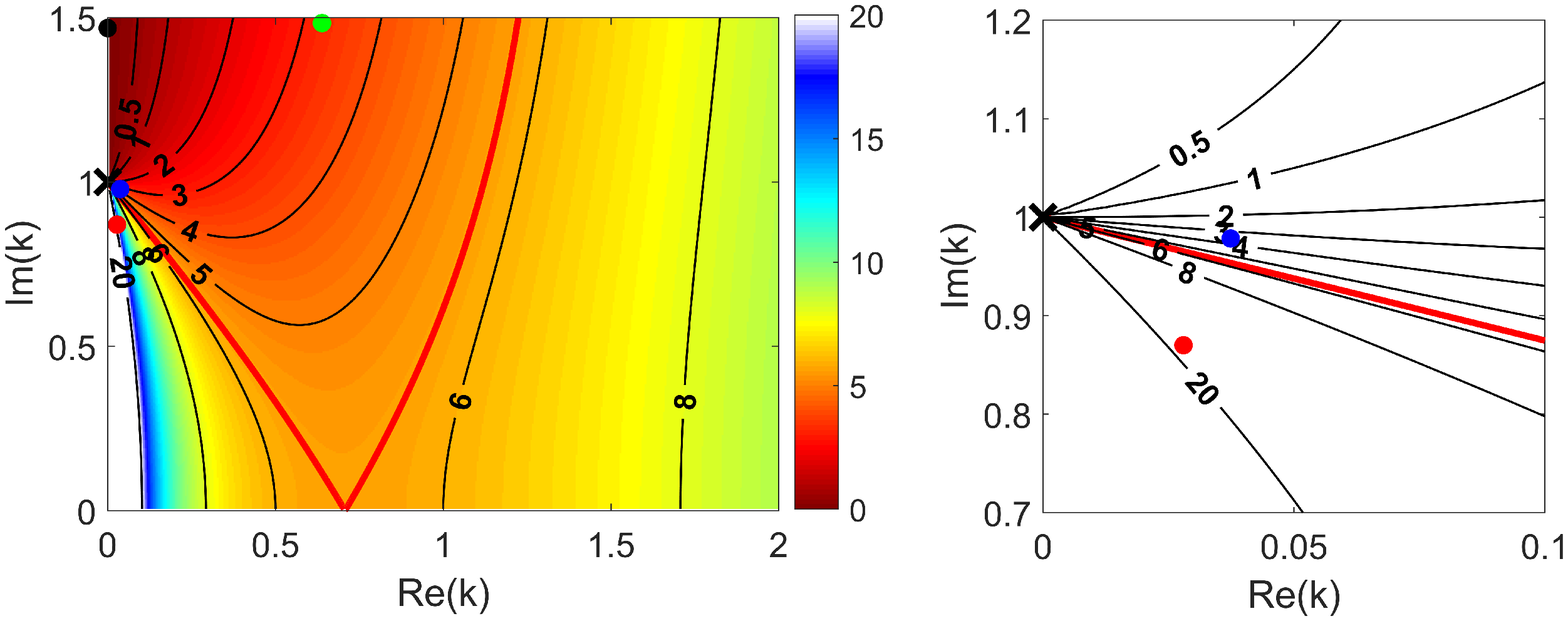}
\caption{False color plot and contour lines of breathers velocity $V_s$ from Eq.~(\ref{vel}) in the right half complex $k$ plane. Red solid line stands for the asymptotic wedge velocity $V_w=4\sqrt{2}=5.66$, $q_0=1$. Right panel shows a zoom around point $k=i$ (cross mark; corresponding to Peregrine rational soliton \cite{Kibler10}).
Dots indicate the breathers obtained via IST for the other figures: 
Red dot ($k= 0.029+0.870i$), SRBs with $V_s>V_w$, Fig.~\ref{fig:SRB_ext}; Blue dot ($k=0.0374+0.9784i$), SRBs with $V_s<V_w$, Fig.~\ref{fig:SRB_int}; Black dot ($k=1.4678i$), KM soliton in Fig.~\ref{fig:KM}(a); Green dot ($k=0.6386+1.483i$), breather pair of Fig.~\ref{fig:KM}(b).
}
\label{fig:velocity}
\end{figure}

In contrast, the same argument applied to a Gaussian envelope leads to the law $g(\omega)z - t^2/t_0^2 = const.$, which implies the peaks to emerge along a parabolic locus in $(t,z)$ plane. The simulation in Fig.~\ref{fig:gaus}(a) shows that this is indeed the case. Note that in Fig.~\ref{fig:gaus}(a) the Gaussian modulation has the same parameters ($\omega$, $a$, and width at half maximum) of the sech-case shown in Fig.~\ref{fig:SRB_ext}(a), and the two input Fourier spectra compared in Fig.~\ref{fig:gaus}(b) are quite similar except for their asymptotic slopes. Remarkably, however, the evolution differs completely from Fig.~\ref{fig:SRB_ext}(a), since in Fig.~\ref{fig:gaus}(a) no breathers emerge (as also confirmed by IST analysis), and the dynamics is asymptotically confined in the wedge. At variance also with the cases shown in Fig.~\ref{fig:wedge} (where $\omega=0$), in this case, the spatio-temporal structure follows the parabolic locus dictated by the MI amplification mechanism, as clearly shown by red dashed line in Fig.~\ref{fig:gaus}(a).

A key point to understand this result  is that the CW background $q_0$ uniquely fixes the asymptotic velocity $V_w$, whereas the velocity $V_s$ of the emerging breathers is also affected by the perturbation parameters.  In order to better understand the interplay between these two velocities, we show in Fig.~\ref{fig:velocity} a level plot of velocity $V_s$ in Eq.~(\ref{vel}) in the complex $k$-plane. We report only the right half semi-plane $\Re(k)>0$, since a mirror symmetry applies for $\Re(k)<0$. Importantly, the contour line corresponding to the wedge velocity $V_w$ (red curve in Fig.~\ref{fig:velocity}) divides the plane into a central domain where breathers are slow ($V_s<V_w$) and two disjoint, left and right, domains where they are fast ($V_s>V_w$). 
In the domain to the right, breathers are not excitable via MI, i.e. with small perturbations of unstable frequencies. Viceversa, the domain to the left contains SRBs that are generated via MI and propagate externally to the wedge, like in the example in Fig.~\ref{fig:SRB_ext} (see the red dot in Fig.~\ref{fig:velocity}). When changing the parameters of the perturbation, the eigenvalue moves and the velocity change. We have found that the approximate relation $V_s \approx g(\omega) t_0$ constitutes a simple rule of thumb to predict the velocity of emergent breathers. Indeed, the velocity turns out to be nearly independent on amplitude $a$, and grows larger by increasing the width $t_0$ or the frequency $\omega$ (up to peak MI gain $\omega=\sqrt{2}$). 
On one hand, there is no upper bound to $V_s$ ($V_s \rightarrow \infty$ as the imaginary axis is approached, which require very large perturbations, i.e. quasi-periodic case). On the other hand, such breathers continue to exist also when crossing into the central region with $V_s<V_w$, and hence could be expected to interact with the auto-modulation, being internal to the wedge. 

\begin{figure}[t!]
\centering
\includegraphics[width=\linewidth]{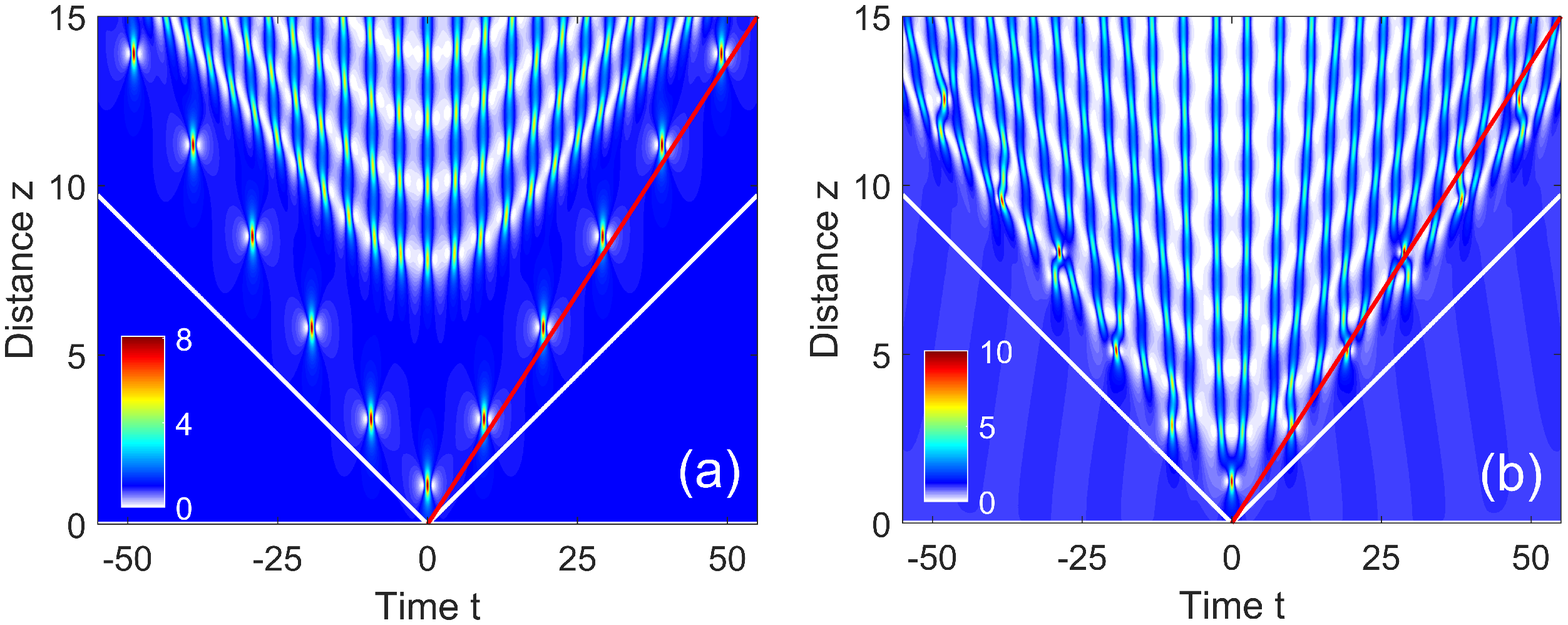}
\caption{Numerical solution of NLSE (\ref{nls}) with initial condition: (a) exact breather pair with eigenvalues $k=\pm 0.0374+0.9784i$ ($R=1.15$, $\alpha=\pm 0.25$);
(b) as in Eq.~(\ref{IV}), sech-shaped, parameters $a=0.543$, $\phi=\pi/2$, $t_0=4$, $\omega=0.545$. In (a,b) $q_0=1$ and solid red line marks the right soliton velocity $V_s=3.668$.}
\label{fig:SRB_int}
\bigskip
\centering
\includegraphics[width=\linewidth]{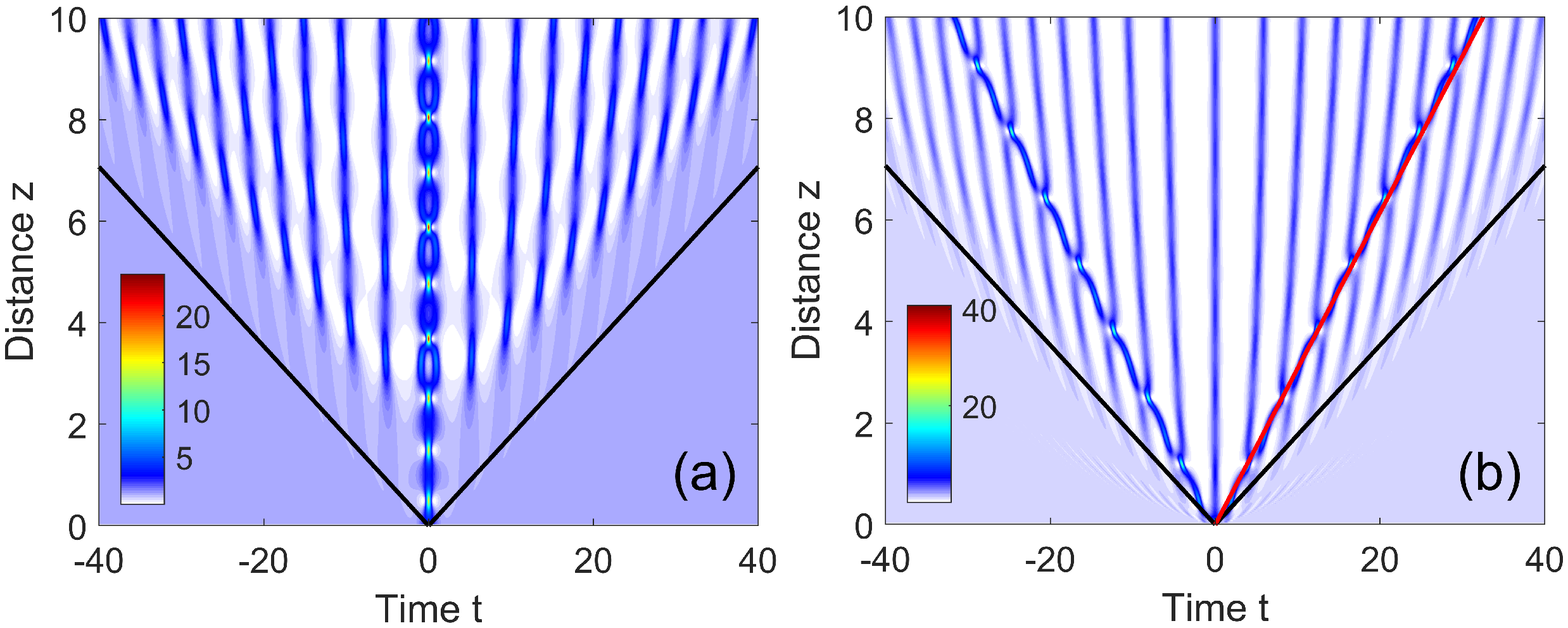}
\caption{ Numerical solution of NLSE (\ref{nls}), input as in Eq.~(\ref{IV}), $q_0=1$: 
(a) Gaussian shape with parameters: $a=0.9$, $\phi=0$, $t_0=1$, $\omega=0$. 
(b) ${\rm sech}$ shape with parameters: $a=5$, $\phi=\pi/2$, $t_0=0.4$, $\omega=4$. 
The IST analysis gives: (a) a single eigenvalue $k=1.468 i$ ($R=2.54$, $\alpha=0$), i.e. a KM soliton with period $z_p=0.998$;
(b) a pair $k=\pm 0.639+1.483i$ ($R=3.035$, $\alpha=0.492$), yielding breather velocities $V_s=\pm 3.252$ (dashed red line).}
\label{fig:KM}
\end{figure}

In Fig.~\ref{fig:SRB_int} we show indeed that SRBs can also be excited inside the wedge. The phenomenon can be more easily recognised by contrasting the case in Fig.~\ref{fig:SRB_int}(a), where we launch the exact breather pair (exactly as in \cite{GZ14,Kibler15}, with parameters $R=1.15$, $\alpha=0.25$, $\mu_{12}=0$, $\theta_{12}=0$) with the case, shown in Fig.~\ref{fig:SRB_int}(b), of a sech-shaped perturbation. It is clearly seen that, in the former case, the propagation is dominated by a breather pair that exhibits long period.
In this case, since the initial condition ideally contains only discrete spectrum, the auto-modulation is seeded only by the numerical error and appears at very large distances. In Fig.~\ref{fig:SRB_int}(b) we tuned the parameter of the sech-shaped perturbation in order to produce the same pair of breathers (as can be verified with our IST analysis, see the blue dot in Fig.~\ref{fig:velocity}). Multiple collisions between the breathers and the modulated structure can be clearly seen. Such collisions are elastic and produce temporal shifts on both the modulation (the central maximum of Fig.~\ref{fig:SRB_int}(a) is no longer present in Fig.~\ref{fig:SRB_int}(b)) and the breathers, which however retain their solitary structure. 

Returning to Fig.~\ref{fig:velocity}, it is clear that the central region of slow breathers include Kuznetsov-Ma (KM) solitons, which stand on the imaginary axis with $\Im(k)>1$, where $V_s \rightarrow 0$ \cite{Kibler12,Dudley14}. In general, KM or quasi-KM pairs with opposite small velocities can be excited with relatively large perturbations. In Fig.~\ref{fig:KM} we illustrate two different examples, showing that  also such breathers undergo interaction with the auto-modulation. In Fig.~\ref{fig:KM}(a) we show the evolution of a positive ($\phi=0$) Gaussian perturbation with $\omega=0$, $a=0.9$, $t_0=1$ for which our IST analysis yields a KM breather with $k=1.4678i$, $V_s=0$ (see also black dot in Fig.~\ref{fig:velocity}). Its signature in Fig.~\ref{fig:KM}(a) is the initial breathing, which, however, soon evolves, due to the interaction with the central peak of the emerging auto-modulation, into periodic cycles of attraction and repulsion (similar to bound state of 2-solitons in the case of zero background). We point out that such dynamics persists also for substantially weaker perturbations, 
though the period rapidly increases due to the shift of the eigenvalue towards $k=1i$ (cross in Fig.~\ref{fig:velocity}), thus making more difficult to recognize the breather signature in the dynamics. 
To excite the non-degenerate case of quasi-KM breather pairs, we need to consider non-vanishing frequencies $\omega$. In the example shown in Fig.~\ref{fig:KM}(b), a sech-shaped envelope modulating the frequency $\omega=4$ gives a pair of breathers. In this regime, the approximate relation $V_s \simeq g(\omega)t_0$ no longer holds valid, and the velocity must be obtained from IST analysis,
which yields eigenvalues $k=\pm 0.6386+1.483i$ (see green dot in Fig.~\ref{fig:velocity}) and in turn, from Eq. (\ref{vel}), a velocity $V_s<V_w$. As a consequence, the pair is observed to collide elastically with the peaks of the auto-modulation inducing mutual temporal shifts at each collision.

In summary, we have illustrated several new scenarios of nonlinear MI which are readily observable in fiber experiments.
We have shown how the control of the perturbation can dramatically change the existence condition for breathers and their interplay with the omnipresent auto-modulation.  

\section*{Funding}
Agence Nationale de la Recherche (ANR) (ANR-11-EQPX-0017, ANR-11-LABX-0007, ANR-14-ACHN-0014); CPER Photonics for Society P4S; IRCICA.
National Science Foundation (DMS-1614623 and DMS-1615524).
S.T. acknowledges kind hospitality from Univ. of Lille, PhLAM.

%
%


\begin{thebibliography}{30}

\bibitem{ZO09} 
V. E. Zakharov and L. A. Ostrovsky, 
Physica D {\bf 238}, 540 (2009).




\bibitem{Kibler10}  
B. Kibler, J. Fatome, C. Finot, G. Millot, F. Dias, G. Genty, N. Akhmediev, and J. M. Dudley, 
Nat. Phys. {\bf 6}, 790 (2010).


\bibitem{Kibler12}  
B. Kibler, J. Fatome, C.Finot, G. Millot, G. Genty, B. Wetzel, N. Akhmediev, F. Dias, and J. Dudley, 
Sci. Rep. {\bf 2}, 463 (2012).


\bibitem{Dudley14} 
J. M. Dudley, F. Dias, M. Erkintalo, and G. Genty, 
Nature Photon. {\bf 8}, 755 (2014).

\bibitem{optimalMI} 
A. Bendhamane, A. Mussot, P. Szriftgiser, A. Kudlinski, M. Conforti, S. Wabnitz, and S. Trillo,  
Opt. Exp. {\bf 23}, 30861 (2015).

\bibitem{Mussot18} A. Mussot, C. Naveau, M. Conforti, A. Kudlinski, F. Copie, P. Szriftgiser, and S. Trillo, 
Nature Photon. {\bf 10}, 303 (2018).



\bibitem{K67} 
V. I. Karpman, 
Pisma Zh. Eksp. Teor. Fiz. {\bf 6}, 829 (1967) [JETP Lett {\bf 6}, 277 (1967)].

\bibitem{KK68} 
V. I. Karpman and E. M. Krushkal, 
Zh. Eksp. Tear. Fiz. {\bf 55}, 530 (1968) [Sov. Phys. JETP {\bf 28}, 277 (1969)].

\bibitem{El93} 
G. El, A. Gurevich, V. Khodorovskii, and A. Krylov, 
Phys. Lett. A {\bf 177}, 357 (1993).

\bibitem{BM16} 
G. Biondini and D. Mantzavinos, 
Phys. Rev. Lett. {\bf 116}, 043902 (2016).

\bibitem{BLM16} 
G. Biondini, S. Li and D. Mantzatinos, 
Phys. Rev. E {\bf 94}, 060201R (2016).

\bibitem{BM17} 
G. Biondini and D. Mantzavinos, 
Commun. Pure Appl. Math. {\bf 70}, 2300 (2017).

\bibitem{BK14} 
G. Biondini and G. Kovacic, 
J. Math. Phys. {\bf 55}, 031506 (2014).

\bibitem{kuznetsov}
E. A. Kuznetsov, 
Sov. Phys. Dokl. \textbf{22}, 507  (1977)

\bibitem{BF15}
G. Biondini and E. R. Fagerstrom, 
SIAM J. Appl. Math. {\bf 75}, 136 (2015).


\bibitem{BLMT17} 
G. Biondini, S. Li, D. Mantzavinos, and S. Trillo, 
arXiv:1710.05068 [nlin.ps] (2017), to appear in SIAM Review


\bibitem{Kraych18} 
A. E. Kraych, P. Suret, G. El, and S. Randoux, 
arXiv:1805.05074 [nlin.ps] (2018).

\bibitem{Audo18} 
F. Audo, B. Kibler, J. Fatome, and C. Finot, 
Opt. Lett. {\bf 43}, 2864 (2018).

\bibitem{ZG13} 
V. E. Zakharov and A. Gelash, 
Phys. Rev. Lett. {\bf 111}, 054101 (2013).

\bibitem{GZ14} 
A. A. Gelash and V. E. Zakharov, 
Nonlinearity {\bf 27}, 1 (2014).

\bibitem{Kibler15} 
B. Kibler, A. Chabchoub, A. Gelash, N. Akhmediev and V. E. Zakharov, 
Phys. Rev. X {\bf 5}, 041026 (2015).

\bibitem{Kibler17} 
B. Kibler, A. Chabchoub, A. Gelash, N. Akhmediev, and V. E. Zakharov, in Nonlinear Guided Wave Optics, S. Wabnitz, Ed. (IOP Publishing, Bristol, UK, 2017), Chap. 7.

\bibitem{Gelash18} 
A. Gelash, 
Phys. Rev. E {\bf 97}, 022208 (2018).

\bibitem{Zhao16} 
L.-C. Zhao and L. Ling, 
J. Opt. Soc. Am. B {\bf 33}, 850 (2016).

\bibitem{Boff1992} 
G. Boffetta and A. Osborne, 
J. Comput. Phys. {\bf 102}, 252 (1992).



\end{thebibliography}



\end{document}